\documentclass[aps,pre,showpacs,reprint,groupedaddress]{revtex4-1}
\usepackage{graphicx}
\usepackage{graphics}
\usepackage{epsfig}
\usepackage{enumerate}

\begin{document}
\title{The dynamics of competitive learning:\\ the role of updates and memory }
\author{Ajaz Ahmad Bhat  }
\email{ajaz@bose.res.in}
\author{Anita Mehta}
\email{anita@bose.res.in}
\affiliation{Theoretical Science Department, S N Bose National Centre, Block JD Sector III, Salt Lake, Kolkata 700098, India}
\date{Revised version: \today}
\begin{abstract}
We examine the effects of memory and different updating paradigms in a game-theoretic model of competitive learning, where agents are influenced in their choice of strategy by both the choices made by, and the consequent success rates of, their immediate neighbours. We apply parallel and sequential updates in all possible combinations to the two competing rules, and find, typically, that the phase diagram of the model consists of a disordered phase separating two ordered phases at coexistence. A major result is that the corresponding critical exponents  belong to the generalised universality class of the voter model. When the two strategies are distinct but not too different, we find the expected linear response behaviour as a function of their difference.
Finally, we look at the extreme situation when a superior strategy, accompanied by a short memory of earlier outcomes, is pitted against its inverse; interestingly, we find that a long memory of earlier outcomes can occasionally compensate for the choice of a globally inferior strategy.
\end{abstract}
\maketitle
\section{Introduction}
\label{intro}
The modelling of social behaviour is of increasing concern to statistical physicists \cite{castellano}. Studies of social and
biological systems often reveal that even when the interactions of a given individual are very localised in time and space,
collective, regular behaviour can emerge: this is analogous to the cooperative behaviour manifested by emergent systems
in the natural world. Such social regularities may well take the form of \textit{learning}, when individuals adopt the behaviour
of other individuals. From the perspective of game theory \cite{games}, this can be seen as an adoption of a particular
\textit{strategy}, whose result may or may not be associated with a favourable outcome. It is then quite reasonable to expect
that the effectiveness of a strategy in yielding favourable outcomes should influence how likely it is to persist, and
spread through the population; the resulting ideas of \textit{strategic learning} \cite{young} have found wide application, starting
from economics \cite{kalyan} to cognitive science \cite{camerer}. 

Against the backdrop of the above ideas, a model of strategic learning was introduced in \cite{mehta}, with one of two possible strategies (denoted
as $+$ and $-$ in the remainder of this paper) being available to each agent on a lattice: the agents were referred to as `myopic'
(aware only of their immediate neighbours) and `memoryless' (unaware of their own and others' past outcomes) in the paper on technology diffusion \cite{kalyan} that inspired the above model \cite{mehta,mahajan}. The question on which this body of work has centred is: despite these handicaps, can agents overall learn to use the superior one of two available technologies? Briefly, each agent changes (or does not change) strategy based on two elementary rules at every time step: a \textit{majority-based} rule, reflecting its tendency to align with its local neighbourhood, followed by a \textit{performance-based} rule, where the agent adopts the strategy that `wins' in its neighbourhood. This (relative) success is measured in terms of \textit{outcomes}, where the probability of a successful outcome for strategy $+$ ($-$) is $p_+$ ($p_-$). Also, the model of \cite{mehta} added to the description of~\cite{kalyan} by endowing the agents with memory:  those agents who make their choices on the basis of the last payoff alone, are adjudged to be memoryless (with a corresponding parameter $\varepsilon$ near $1$), while those who allow for memories of earlier outcomes may make decisions
that run counter to immediate evidence ($\varepsilon$ small).  

Some related ideas have been examined in recent work. For example, the issue of consensus formation in a model of threshold learning \cite{vega} shows close analogies: in this model, the competition between the `noisy' signals from the immediate neighbours of an agent (cf. the majority rule in \cite{mehta}) and the 
 acceptance threshold that agents require to change their state (cf. the memory threshold in the performance-based rule of \cite{mehta}), determine 
the phase diagrams obtained.  Recent studies of coevolving Glauber dynamics on networks \cite{mandra} are also relevant, since the model of \cite{mehta} can be viewed as a competition between the Glauber dynamics of two sets of Ising spins, corresponding to strategy and outcome respectively.

In the current paper, we take all these ideas further. First, we explore the effect of different updates. If new information propagates \textit{sequentially} through the network, and the arrow of time is discernible in the decisions of individual agents, are the global phase diagrams any different from what they would be if information was transmitted and all decisions were taken \textit{simultaneously}? Common sense tells us that sequential or parallel updates \textit{should} make a difference to the nature of the phase diagram, and the results of the present paper confirm this. Also  (unlike the work of \cite{mehta, mahajan} which examined the situation at coexistence) we look in this paper at the effects of disparate strategies ($p_+ \neq p_-$). The final, and possibly most important issue, is that of memory, which acts as a threshold governing change \cite{vega}: what is the effect of the threshold $\varepsilon$, which tells the agent that longer-term inputs are significant, and need to be considered when making a decision? We will find that, indeed, a longer memory of earlier outcomes can sometimes make up for the choice of a globally inferior strategy.

The plan of this paper is as follows. In Section~\ref{defsection}, we review the model of \cite{mehta}. In Section~\ref{updates}, we discuss the behaviour of the model for a range of updating schemes, in the presence of memory. In Section~\ref{away}, we examine the behaviour of the model away from coexistence, as a function of \textit{distinct} parameter values for the two strategies; in particular we discuss here the role of memory. In the concluding section, we discuss our results and put them in the context of other recent work in the field.
%%%%%%%%%%%%%%%%%%%%%%%%%%%%%%%%%%%%%%%%%%%%%%%%%%%%%%
%%%%%%%%%%%%%%%%%%%%%%%%%%%%%%%%%%%%%%%%%%%%%%%%%%%%
\section{Definition of the model}
\label{defsection}
The model of \cite{mehta} involves two types of strategies,  $-$ and $+$,   where the $+$ strategy is globally superior \cite{kalyan} to the $-$ strategy.  As mentioned above, agents tend to follow the strategy adopted by the majority of their neighbours, modifying this choice in a second step (if necessary)
according to which of these have proved to be the most successful.

Assuming that the agents sit at the nodes of a \textit{d}-dimensional regular lattice with coordination number $z = 2d$, the efficiency of an agent at site \textit{i} is represented by an Ising spin variable:
\begin{equation}\eta_{i}(t) = \left\{\begin{array}{ll}
+1 & \mbox{ if \textit{i} is $+$ at time \textit{t},} \\
-  1 & \mbox{ if \textit{i} is $-$ at time \textit{t}.}
\end{array}
\right.\end{equation}
The evolution dynamics of the lattice is
 governed by two rules. The first is a \textit{majority} rule, which consists of the alignment of an agent with the local field (created by its nearest neighbours) acting upon it, according to:
\begin{equation}
\eta_{i}(t+\tau_{1})=\left\lbrace\begin{array}{cccc}+1 &  & \mbox{if}\ & h_{i}(t)> 0,\\\pm 1  & \makebox[1.5cm]{w.p}\frac{1}{2} & \mbox{if}\ & h_{i}(t)=0,\\-1 &    &  \mbox{if}\ & h_{i}(t)<0. \end{array} \right .
\end{equation}
Here, the local field
\begin{equation}
\mbox{$h_\textit{i}(t)$ = $\sum_{\textit{j(i)}}{\eta_\textit{j}(t),}$}
\end{equation}
is the sum of the efficiencies of the $\textit{z}$ neighbouring agents $\textit{j}$ of site $\textit{i}$ and $ \tau_{1}$ is the associated time step.
Next, a \textit{performance} rule is applied. This starts with the assignment of an outcome $\sigma_{i}$ (another Ising-like variable, with values of $\pm 1$ corresponding to success and failure respectively) to each  site $\textit{i}$, according to the following rules:
\begin{eqnarray}
\mbox{if}\  \eta_{i}(t)= +1,\nonumber\\
 \mbox{then } & \sigma_{i}(t+\tau_{2})=\left\lbrace\begin{array}{ccc}+1 & \mbox{w.p.} & p_{+}\\-1\  & \mbox{w.p.}\  & 1- p_{+},\end{array}\right .\nonumber\\
\mbox{if}\  \eta_{i}(t)= -1,\nonumber\\
 \mbox{then } & \sigma_{i}(t+\tau_{2})=\left\lbrace\begin{array}{ccc}+1 & \mbox{w.p.} & p_{-}\\-1\  & \mbox{w.p.}\  & 1- p_{-},\end{array}\right .
\end{eqnarray}
where $\tau_{2}$ is the associated time step and $p_\pm$ are the probabilities of having a successful outcome
for the corresponding strategy. With $N_{i}^{+}$ and $N_{i}^{-}$  denoting  the total number of neighbours of a site \textit{i} who have adopted strategies $+$  and $-$  respectively, and  $I_{i}^{+}$ ($I_{i}^{-}$) denoting the number of successful outcomes within the set  $N_{i}^{+}$ ($N_{i}^{-}$), the dynamical rules for site \textit{i} are:
\begin{eqnarray}
\mbox{if}\ \eta_{i}(t)=  +1  & \mbox{ and}\ \frac{I_{i}^{+}(t)}{N_{i}^{+}(t)}<\frac{I_{i}^{-}(t)}{N_{i}^{-}(t)},\nonumber\\
\mbox{then } & \eta_{i}(t+\tau_{3})=\left\lbrace\begin{array}{ccc}-1 & \mbox{w.p.} & \varepsilon_{+}\\+1 & \mbox{w.p.} & 1- \varepsilon_{+},\end{array}\right .\nonumber\\
\mbox{if}\ \eta_{i}(t)=  -1  & \mbox{ and}\ \frac{I_{i}^{-}(t)}{N_{i}^{-}(t)}<\frac{I_{i}^{+}(t)}{N_{i}^{+}(t)},\nonumber\\
\mbox{then } & \eta_{i}(t+\tau_{3})=\left\lbrace\begin{array}{ccc} +1 & \mbox{w.p.} & \varepsilon_{-}\\ -1 & \mbox{w.p.} & 1- \varepsilon_{-}.\end{array}\right .
\end{eqnarray}

Here, the ratios $\frac{I_{i}(t)}{N_{i}(t)}$ are nothing but the average payoff assigned by an agent to each of the two strategies in its neighbourhood at time $t$ (assuming that success yields a payoff of unity and failure, zero). Also, $\tau_{3}$ is the associated time step and the parameters $\varepsilon_{\pm}$ are indicators of the memory associated with each strategy. In their full generality, $\varepsilon$ and $p$ are independent variables: the choice of a particular strategy can be associated with either a short or a long memory. However, we would like in this paper to answer a question which was posed, but not answered in \cite{mehta}: can the presence of a good memory compensate for the choice of an inferior strategy? 
We therefore examine the extreme situation when a
 globally superior strategy ($p_{+} \gg p_- $), combined with a shorter memory ($\varepsilon_{+} \gg \varepsilon_{-}$) is in competition with its inverse: this is the situation  that will be studied in Section~\ref{away}.

 Setting the timescales 
\begin{equation}
\tau_{2} \rightarrow 0,\  \tau_{1} = \tau_{3} = 1, 
\end{equation}
the above steps of the performance rule are recast  as effective dynamical rules involving the efficiencies $\eta_{\it{i}}(t)$ and the associated local fields alone:
\begin{eqnarray}
\mbox{if}\  \eta_{i}(t)= +1,\nonumber\\
\mbox{then}\  \eta_{i}(t+1) &= \left\lbrace\begin{array}{ccc}+1\  & \mbox{w.p.} &\mbox{$w_+$[$h_{i}(t)$]}\\-1\  & \mbox{w.p.}\  &\mbox{1-$w_+$[$h_{i}(t)$]},\end{array}\right .\nonumber\\
\mbox{if}\  \eta_{i}(t)= -1,\nonumber\\
\mbox{then}\  \eta_{i}(t+1) &= \left\lbrace\begin{array}{ccc}+1\  & \mbox{w.p.} &\mbox{$w_-$[$h_{i}(t)$]}\\-1\  & \mbox{w.p.}\  &\mbox{1-$w_-$[$h_{i}(t)$]}.\end{array}\right .
\end{eqnarray}
The effective transition probabilities $w_{\pm}(h)$ are evaluated by enumerating the $2^{z}$ possible realizations of the outcomes $\sigma_{j}$ of the sites neighbouring site \textit{i}, and weighing them appropriately. For a 2-$d$ square lattice, the possible local field values at the interfacial sites are 0 and $\pm$2. The corresponding transition probabilities for these field values are \cite{mehta}:
\begin{eqnarray}
w_{+}(+2) = 1 - \varepsilon_{+}p_{-}(1-p_{+}^3),\nonumber\\
w_{-}(+2) = \varepsilon_{-}(1 - p_{-})[1 -(1-p_{+})^3],\nonumber\\
w_{+}( 0) = 1 - \varepsilon_{+}p_{-}(1 - p_{+})(2 - p_{-} - 2p_{+} + 3p_{-}p_{+}),\nonumber\\
w_{-}( 0) = \varepsilon_{-}p_{+}(1 - p_{-})(2 - p_{+} - 2p_{-} + 3p_{-}p_{+}),\nonumber\\
w_{+}(-2) = 1 - \varepsilon_{+}(1 - p_{+})[1 -(1-p_{-})^3],\nonumber\\
w_{-}(-2) = \varepsilon_{-}p_{+}(1-p_{-}^3)\nonumber.\\
\label{teqns}
\end{eqnarray}
    In \cite{mehta},  the model was explored at coexistence with an ordered sequential update applied to memoryless agents \cite{kalyan}:
\begin{equation}
p_{+} = p_{-}, \hspace*{1 cm}  \varepsilon_{+} = \varepsilon_{-} = 1.
\end{equation}
In the present paper, we go beyond this in two different ways. First, still at coexistence, we explore the effect of different updates on the $p-\varepsilon$ phase diagram of the model: next, we examine the model away from coexistence, for distinct values of 
$p_{\pm}$ and $\varepsilon_\pm$. The basic quantities considered hereafter are the magnetization $M$, staggered magnetization $M_{stag}$ and the energy $E$.  These quantities are defined for a finite sample of $N$ agents (or sites) and $N$\textit{z}/2 bonds (or links), as\\

\begin{eqnarray}
 M=\frac{1}{N}\sum_{i}\eta_{i}\ ,\nonumber \\
E=\frac{1}{N\textit{z}}\sum_{ij}(1-\eta_{i}\eta_{j})\ ,\nonumber \\
 M_{stag}= \frac{2}{N}\sum_{i}\eta_{i} \;    \mbox{if \textit{i} is odd or even.}
\label{defs}
\end{eqnarray}
In the following we shall usually consider mean values  \(\langle M \rangle\),   \(\langle\)E\(\rangle\) and  \(\langle M_{stag}\rangle\).
% % % % % % % % % % % % % % % % % % % % % % % % % % % % % % % % % % % % % %
% % % % % % % % % % % % % % % % % % % % % % % % % % % % % % % % % % % % % %
\section{The effect of finite memory, and of different updates}
\label{updates}
We begin this section with a review of the physical significance of updating schemes. Most generally,  updates can be random or ordered as follows:
     \begin{itemize}
 \item \textit{Random}: Here,  sites are chosen at random for the consecutive application of rules.
		\item \textit{Ordered}: Here, sites are chosen in an ordered fashion, i.e., after choosing every $(i,\ j)^{th}$ site, the  $(i,\ j+1)^{th}$ site is selected.  
		\end{itemize}
Since the sociological basis for this work was the propagation of innovation through connected societies \cite{kalyan}, we choose to deal only with ordered	updates here. However, even ordered updates have two subclasses: parallel and sequential. Assume a condition \textit{A} such that when an agent satisfies \textit{A}, it changes strategy:
\begin{itemize}
\item \textit{Sequential update}: In this type of update, we check the condition \textit{A} on the $(i,\ j)^{th}$ site, then update the efficiency of the site  and proceed to the  $(i,\ j+1)^{th}$ site using the updated value of  the $(i,\ j)^{th}$ site.
\item \textit {Parallel update}: In  this type of update, we check the condition A on the $(i,\ j)^{th}$ site, do \textit{not} update the site but instead save the update-decision in memory, and proceed to the next site. Once the whole lattice is swept, all the saved update-decisions are implemented `simultaneously'.
\end{itemize}
The choice of different updates generally corresponds to different physical situations: it has been shown that it also leads to a disparity in the convergence time of the systems concerned \cite{kanter1,kanter2}. We therefore examine all possible combinations for our two update rules:
\begin{enumerate}[I]
\item {parallel updates for both majority rule and performance rules }(\textit{pp}).
\item {parallel update for majority rule and sequential update for performance rules }(\textit{ps}).
\item{sequential updates for both majority rule and performance rules }(\textit{ss}).
\item {sequential update for majority rule and parallel update for performance rules }(\textit{sp}).
\end{enumerate}

In the following subsections, we explore the phase dynamics at coexistence for each of these
update rules in turn, for both parameters $p$ and $\varepsilon$. We state at the outset that all the updates (except for the \textit{sp} update) which we consider, result in models which are in the general university class of the voter model \cite{haye}: the inverse energy 1/\textit{E(t)} is thus always proportional to the logarithm of time, ln $t$. When, as in the case of the $ss$ update, the value of the slope is exactly  $2/\pi$ \cite{mehta}, the \textit{exact} universality class of the voter model is retrieved.
 % % % % % % % % % % % % % % % % % % % % % % % % %
%%%%%%%%%%%%%%%%%%%%%%%%%%%%%%%%%%%%%%%%%%%%%%%%%%
\subsection{The ss update}
\label{sssection}
\begin{figure}[h]
\includegraphics[scale =0.37]{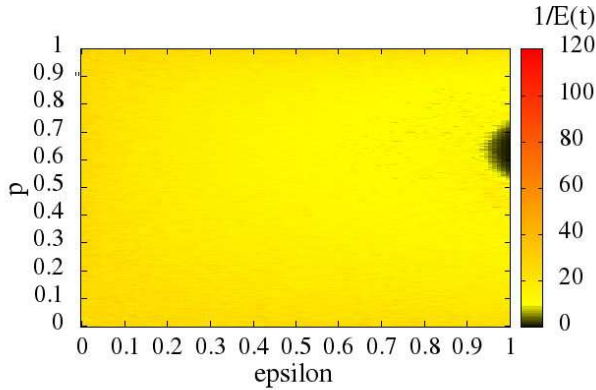} 
  \caption{(color online)[\textit{ss} update] Phase diagram of the model with an $ss$ update. Plot of the inverse energy 1/$E(t)$ at time $t = 512$ for a square lattice of size $N = 64^2$ in the $p$-$\varepsilon$ plane. The black region shows the disordered phase and the yellowish (light grey) region shows the frozen phase.}
  \label{fig:sspd}
\end{figure}

This is the update that was used throughout \cite{mehta}; however the phase behaviour of the
model was there only explored  for the parameter $p$, whereas here we extend it to the parameter
$\varepsilon$. In Figure~\ref{fig:sspd}, we  plot the inverse energy 1/$E(t)$ in the $p-\varepsilon$ plane at time $t = 512$ for a square lattice of size $N = 64^2$.
This phase diagram shows clearly the existence of a disordered paramagnetic phase embedded in a largely frozen phase elsewhere.  The disordered phase exists for \linebreak[4]$p_{c1}(=0.56\pm 0.01)<p<p_{c2}(=0.70\pm 0.01)$ when $\varepsilon \geq 0.980$. Our results agree with those of \cite{mehta} for $\varepsilon = 1$, and extend them all across the rest of the $p-\varepsilon$ plane. We mention here that the average time required to reach consensus increases exponentially as \textit{p} decreases in the frozen phase, leading to the presence
of striped states \cite{redner} at limiting values of $p$. Figure~\ref{fig:sspd} also makes it clear that the effect
of increasing memory wipes out the disordered phase: this is as it should be, since the disordered phase is generated by the competition between the majority and performance-based rules, which is dulled by increasing memory.

\begin{figure}[h]
\centering\resizebox{1.40in}{!}{\includegraphics[0in,0in][4in,3in]{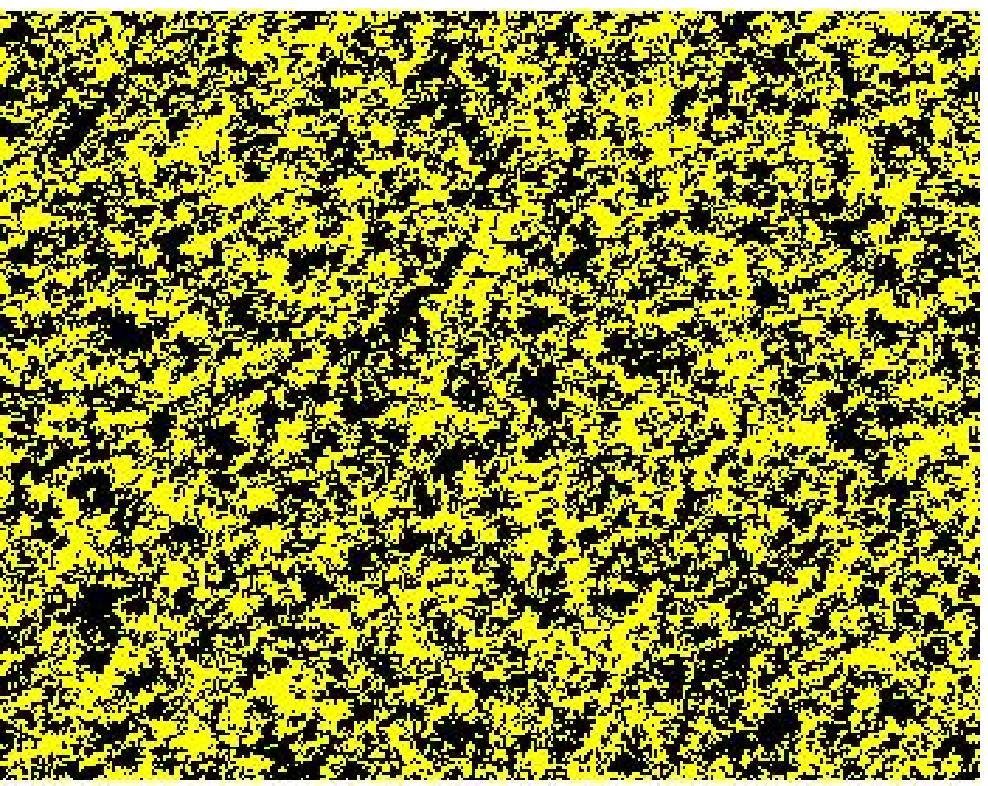}}
\centering\resizebox{1.4in}{!}{\includegraphics[0in,0in][4in,3in]{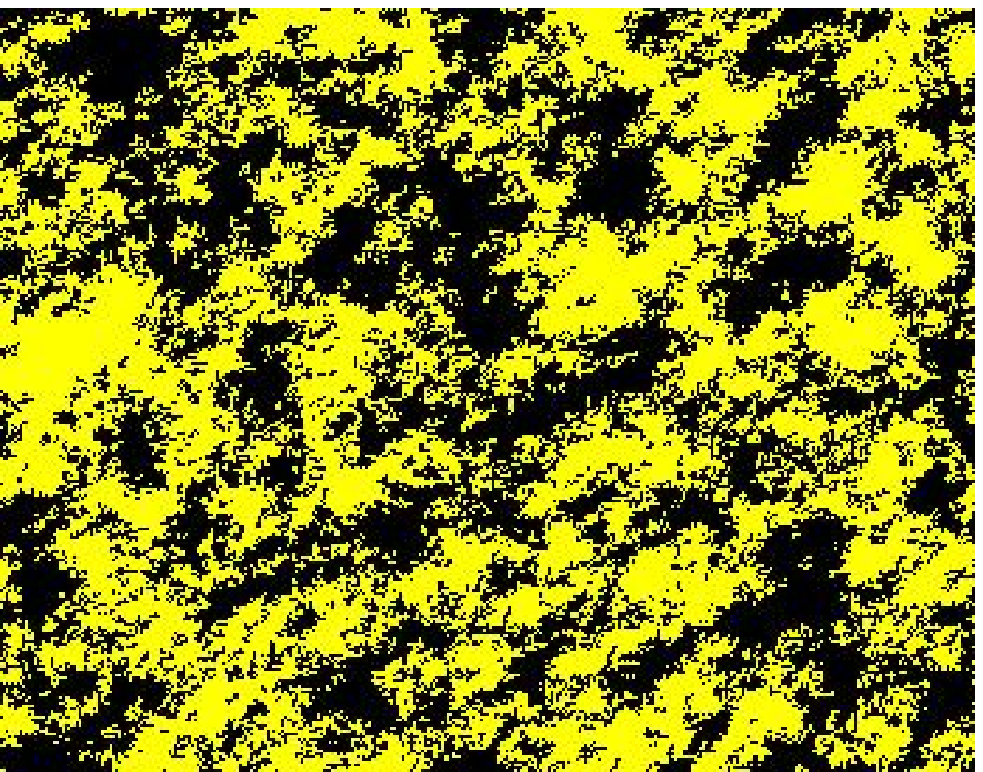}}
\centering\resizebox{1.4in}{!}{\includegraphics[0in,0in][4in,3in]{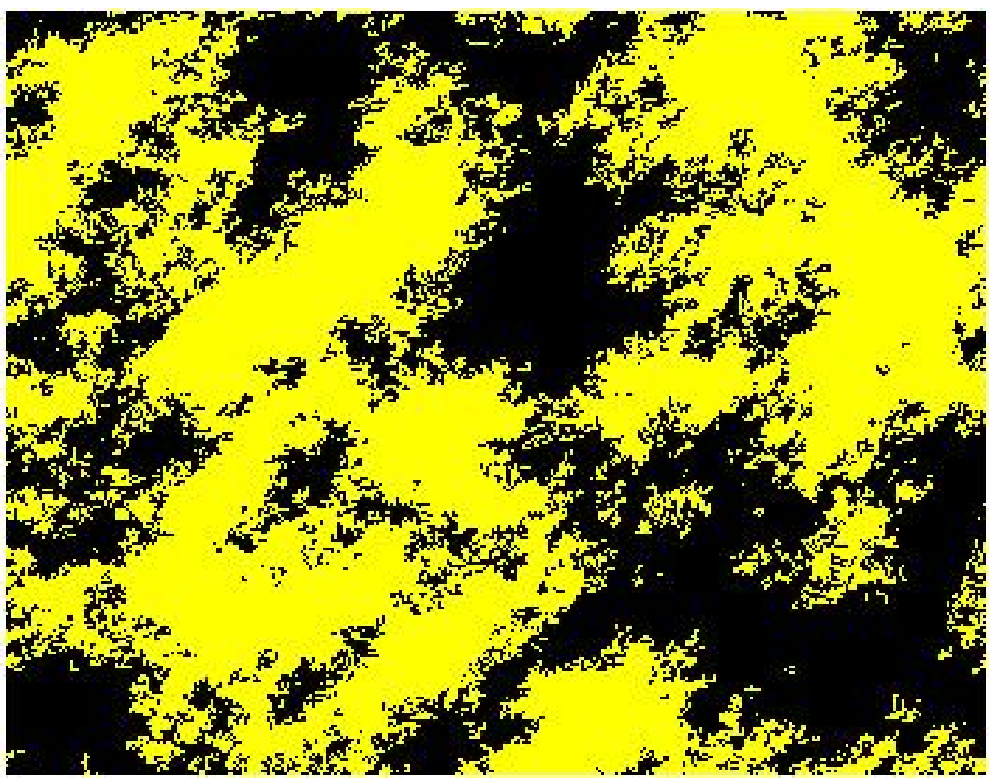}}
\caption{(color online)[\textit{ss} update] Snapshots of the dynamics of the $ss$-updated model. Each plot is a portion (of size $100^2$) of a square lattice of $N = 256^2$ for $p = 0.72$ and $\varepsilon = 1.0$
at times $t = 8$ (top-left), $t = 64$ (top-right) and $t = 512$ (bottom).}
\label{fig:sssnap}
\end{figure}

Figure~\ref{fig:sssnap} shows snapshots of the dynamics of the model using a lattice of size $N = 512^2$ at times $t = 8$, $t = 64$, and $t = 512$ with random initial configurations and parameter values $p=0.72$ (very close to the critical point $p_{c2}$) and  $\varepsilon = 1.0$. 
 The plots reveal characteristically voter-like \cite{haye} coarsening  behaviour.
 
\begin{figure}[h]
\vspace*{0.45cm}
\centering{\includegraphics[scale = 0.28]{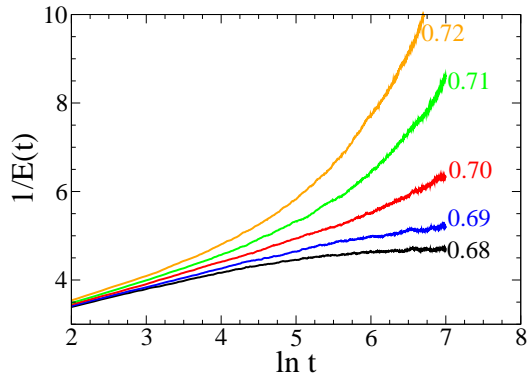} } 
\caption{(color online)[\textit{ss} update] Plot of the inverse energy $1/E(t)$ versus the natural logarithm of time ln $t$ for different values of $p$ close to $p_{c2} = 0.70$,  with  $\varepsilon$ set to 1.0. The lattice size $N = 256^2$, and the $p$-values are indicated on the curves. The curve corresponding to $p = p_{c2} = 0.70$ (shown in red) has slope $2/\pi$ approximately [see Equation~\ref{eqssv}].}
\label{fig:ssv}
\end{figure}

  In Figure~\ref{fig:ssv}, we have plotted the inverse energy 1/\textit{E(t)} against the natural logarithm of time ln $t$ for values of \textit{p} around the critical point $p_{c2} = 0.70 \pm 0.01$. Each of the curves is obtained by averaging over 200 independent samples of size $256^{2}$. At the critical point, we obtain a straight  line with a slope close to $2/\pi$ \cite{mehta}, a behaviour characteristic of the exact voter model~\cite{haye} that corresponds to
\begin{equation}
E(t) \approx \frac{\pi/2}{\mbox{ln}\ t}.
\label{eqssv}
\end{equation}
Similar behaviour is obtained at the other critical point $p_{c1}$, in agreement with \cite{mehta}.
%%%%%%%%%%%%%%%%%%%%%%%%%%%%%%%%%%%%%%%%%%%%%%%%
%%%%%%%%%%%%%%%%%%%%%%%%%%%%%%%%%%%%%%% %%%%%%%%% 
 \subsection{The pp update}
 \label{ppsection}
In this case, both environmental majority and performance-based rules are applied using parallel updates. As we will see, although the universality class of the model is qualitatively unchanged, this update results in the appearance of novel ordered phases compared to the \textit{ss} update. As before, we first plot the phase diagram for all values of $p$ and $\varepsilon$, then show snapshots of the dynamics, and finally get a more quantitative feel for the behaviour of key quantities as a function of $p$.
  
\begin{figure}[h]
   \includegraphics[scale = 0.37]{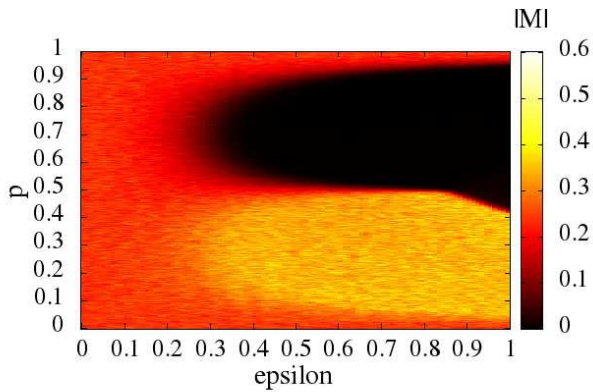} 
   \includegraphics[scale = 0.37]{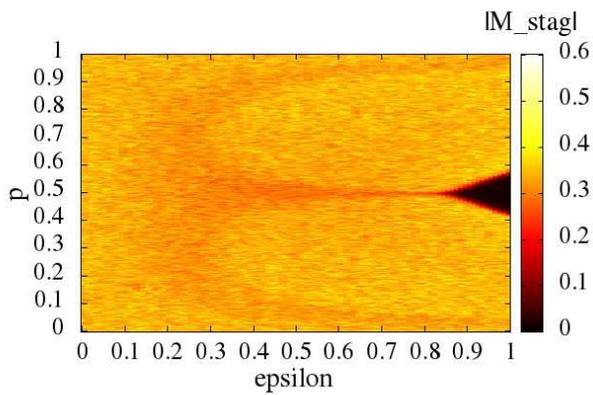} 
   \caption{(color online)[\textit{pp} update] Plot of the absolute value of the magnetization $|M|$ (top) and the absolute value of the staggered magnetization $|M_{stag}|$ (bottom) at time $t = 512$ for a lattice of $N = 100^2$ in the $p$-$\varepsilon$ plane. In the top figure, the yellowish (light grey) region refers to the parallel frozen phase (PFP), while  the black region includes both the anti-parallel frozen phase (AFP), and the disordered region. In the bottom figure, the black region represents the disordered phase characterised by very low $|M_{stag}|$.}
   \label{fig:pppd}
   \end{figure}

 Accordingly, Figure~\ref{fig:pppd} (top and bottom), are  plots of the absolute values of magnetization $|M|$ and staggered magnetization $|M_{stag}|$, at time $t = 512$ for a lattice size $N = 100^2$, in the $p$-$\varepsilon$ plane using \textit{pp} updates. In these phase diagrams, we see clear evidence of the existence of two distinct frozen phases separated  by a disordered phase. Looking along the line $\varepsilon$ = 1, disorder prevails for $p_{c1}<p<p_{c2}$ with $p_{c1}=0.43\pm 0.01$ and $p_{c2}=0.57\pm 0.01$. Notice the symmetry of the two critical points about $p = 0.5$: we shall have more to say about this later on.

   For \textit{p} below $p_{c1}$, there is a frozen phase characterised by overall alignment of spins: we call this the parallel frozen phase (PFP). For \textit{p} above $p_{c2}$, the frozen phase that appears is characterised by an anti-parallel ordering of spins: we call this the anti-parallel frozen phase (AFP).   We  mention also that in the AFP, the lattice may have more than one anti-parallel domain, with thin frustrated  chains running in between them. This frustration can be attributed to the inability of the different domains to align with each other under periodic boundary conditions. The disturbances caused by these chains (in  quantities such as $|M|$ or $E$) due to misalignment decrease as $1/{\sqrt{N}}$ and also appear to vanish for large times. Again, we notice that the phase transition disappears for low  $\varepsilon$; in fact, at very low values of $\varepsilon$ the evolving lattice may get trapped into striped states \cite{redner} at long times.
       \begin{figure}[h]
       \vspace*{0.5cm}
       \hspace*{-7cm}\resizebox{0.65 in}{0.9 in}{\includegraphics[0in,0in][3in,4in]{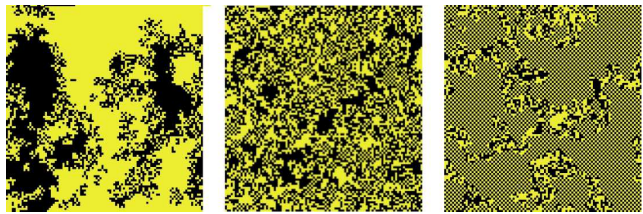}}
       \caption{(color online)[\textit{pp} update] Snapshots of the dynamics of the pp-updated model on a square lattice, for $p = 0.41$ (leftmost), $p = 0.50$ (centre) and $p = 0.59$ (rightmost) at time $t = 512$, with  $\varepsilon$ = 1.0.
        The yellow (light grey) and black colours represent the two strategies, while the greyish grid corresponds to anti-parallel arrangements of yellow (light grey) and black. The leftmost picture represents the PFP (see text), the centre one the disordered phase, and the rightmost one the AFP (see text).}
       \label{fig:ppmat}
       \end{figure}

    This can be understood as follows: the effect of a long memory ($\varepsilon$ small) strongly reduces the relative impact of the performance-based rule. Depending on the value of $\varepsilon$, the performance rule may not be effective for several timesteps whereas the majority rule is implemented at every timestep. In the limit of vanishing $\varepsilon$, then, only the (zero-temperature) majority rule will be effective, leading to stripe formation as predicted by~\cite{redner} for this situation. 
   
 Figure~\ref{fig:ppmat} comprises snapshots of the dynamics of the model for a $2d$ square lattice of size $N = 512^2$ and at time $t = 256$, with random initial configurations. The plots show  a  portion of size $100^2$  of the square lattice for three values of $p$:  $p = 0.41$ (near the critical point $p_{c1}$ between the PFP and the paramagnetic phase), $p = 0.50$ (within the paramagnetic phase) and $p = 0.59$ (near the critical point $p_{c2}$ separating the paramagnetic phase from the AFP), with $\varepsilon = 1$. The snapshot at $p = 0.41$ shows the lattice evolving towards consensus (parallel alignment) with the formation of domains of one type only. The snapshot at $p = 0.50$ shows the lattice in its disordered phase, while the one at $p = 0.59$ shows that the nature of the lattice ordering is anti-parallel. 

\begin{figure}[h]
\vspace*{0.5cm}
\centering\includegraphics[scale =0.28]{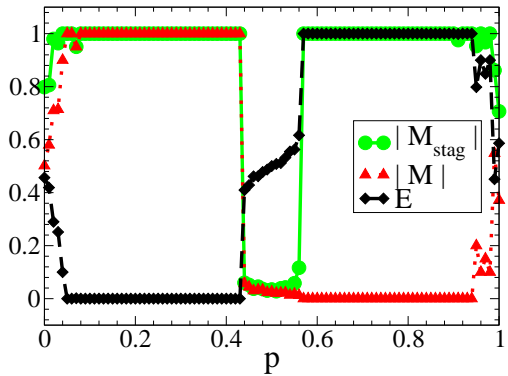} 
\caption{(color online)[\textit{pp} update] Plot of  the absolute value of  magnetization $|M|$, the absolute value of  staggered magnetization $|M_{stag}|$ and energy $E$ against $p$, with  $\varepsilon = 1$ and lattice size $N = 80^2$. Each curve is drawn using symbols (and colour) as indicated in the legend.}
\label{fig:ppmag}
\end{figure}

To investigate this more quantitatively, we plot  the absolute value of  magnetization $|M|$, the absolute value of  staggered magnetization $|M_{stag}|$ and energy $E(t)$ against \textit{p}, with $\varepsilon = 1.0$,  in Figure~\ref{fig:ppmag}. These measurements were recorded  using a square lattice of size $N = 80^{2}$ at time $t = 10^{6}$. All the curves are averaged over 100 independent samples for each value of \textit{p}.  In the region $p \le p_{c1}$,  the values of magnetization $|M|$ and staggered magnetization $|M_{stag}|$ are both equal to unity at saturation,  implying a parallel alignment of the sites; whereas for
 \textit{p} above $p_{c2}$,
the magnetization $|M|$ is zero and the staggered magnetization $|M_{stag}|$ equals unity at saturation, indicating  an anti-parallel alignment of the sites. The energy graph is consistent with this interpretation, given the definition of the energy in Equation~\ref{defs}: zero in the PFP, middling in the paramagnetic phase and unity in the AFP.

\begin{figure}[h]
\vspace*{0.4cm}
\centering\includegraphics[scale =0.28]{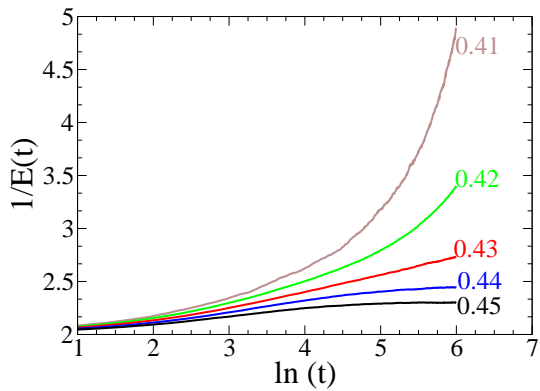}
\caption{(color online)[\textit{pp} update] Plot of   the inverse energy $1/E(t)$ versus  ln $t$ for different values of $p$ close to $p_{c1} = 0.43$ and $\varepsilon = 1$, for a lattice of size $N = 100^2$. The $p$-values are indicated on the curves. The straight line corresponding to $p_{c1} = 0.43$ (shown in red) has slope $1/2\pi$ approximately.}
\label{fig:ppv1}
\end{figure}

\begin{figure}[h]
\vspace*{0.4cm}
\centering\includegraphics[scale = 0.28]{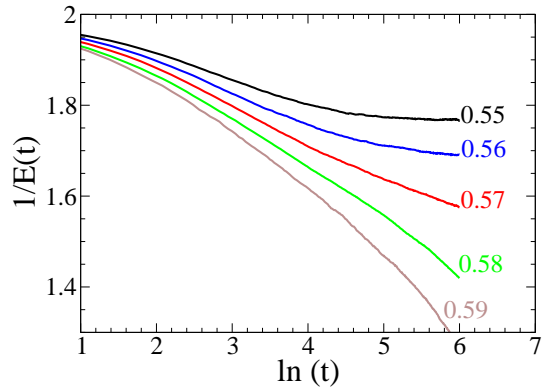}  
\caption{(color online)[\textit{pp} update] Plot of   the inverse energy $1/E(t)$ versus ln $t$ for different values of $p$ close to $p_{c2} = 0.57$,   with  $\varepsilon = 1$, for a lattice of size $N = 100^2$. The $p$-values are indicated on the curves. The straight line corresponding to $p_{c2} = 0.57$ (shown in red) has slope $- 1/5\pi$ approximately.}
\label{fig:ppv2}
\end{figure}

In order to confirm the voter-like nature of the critical points, we plot  the inverse energy $1/E(t)$  against the natural logarithm of time ln $t$, choosing $p$ values near both critical points (see Figure~\ref{fig:ppv1} and Figure~\ref{fig:ppv2}). Each curve is an average over 200 independent samples. Exactly at the critical points $p_{c1} = 0.43$ and $p_{c2} = 0.57$, a linear behaviour of inverse energy with respect to ln $t$ is found, with slopes of $1/2\pi$ and $- 1/5\pi$ respectively. While the critical exponents are those of the voter model \cite{haye}, the  values of the slope are different from $2/\pi$: we find therefore that the \textit{pp} update of the model belongs to the  universality class of the generalised, rather than the exact, voter model \cite{haye}.

To conclude this subsection: the main effect of the \textit{pp} update is to change the nature of the ordering in  one of the two frozen phases, so that anti-parallel ordering is found in the high-$p$ frozen phase. As before, the effect of increasing memory (going to low $\varepsilon$) is to smear out the phase transitions to the disordered phase, by undermining the effect of the outcome-based rule whose competition with the majority rule causes the appearance of disorder. Such instances of mixed domains have been found in recent work on coevolving (parallel) dynamics \cite{mandra}; some features
of these results also appear in studies of threshold dynamics of societal systems \cite{vega}. For a real-life example of the AFP in the case of technology diffusion, we cite the results of  \cite{zhao} where the authors conclude that ``in technology clusters where direct competitors are right next door, leading firms generate innovations that are technologically very distant from their neighbours" \cite{zhao}.
%%%%%%%%%%%%%%%%%%%%%%%%%%%%%%%%%%%%%%%%%%%%%%%%%%%
%%%%%%%%%%%%%%%%%%%%%%%%%%%%%%%%%%%%%%%%%%%%%%%%%%%
\subsection{The ps update}
\label{pssection}
The behaviour of the  \textit{ps}-updated model is qualitatively similar to that of the  \textit{pp}-updated model above. Again, there are  two frozen phases PFP and AFP, separated by a disordered phase: the values of the critical points $p_{c1}$ and $p_{c2}$ are however shifted, such that the disordered region extends between $p_{c1}= 0.31\pm 0.01$ and $p_{c2}= 0.69\pm 0.01$ at $\varepsilon = 1.0$. We find once again that the two critical points are symmetrically placed with respect to $p=0.5$, as in the $pp$ update: we will give an argument for why this is so, in the following subsection.
  \begin{figure}[h]
  \centering\includegraphics[scale = 0.37]{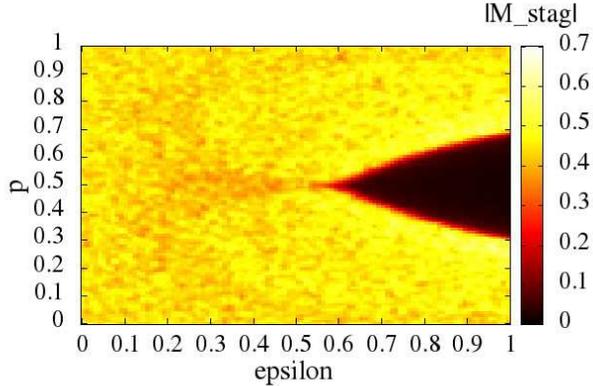} 
  \caption{(color online)[\textit{ps} update] Phase diagram in the $p$-$\varepsilon$ plane of the \textit{ps}-updated model, with a plot of the absolute value of staggered magnetization $|M_{stag}|$ at time $t = 512$, for a  lattice size of $N = 100^2$. The black region represents the disordered phase (very low $|M_{stag}|$), while the yellowish (light grey) region represents frozen phases with high $|M_{stag}|$.}
  \label{fig:pspd}
  \end{figure}

\begin{figure}[h]
 \centering\includegraphics[scale = 0.28]{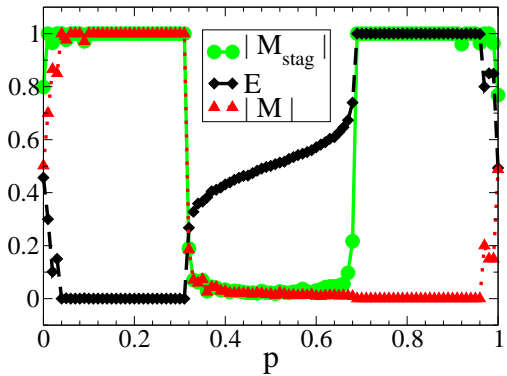}
 \caption{(color online)[\textit{ps} update] Plot of  the absolute value of  magnetization $|M|$, energy $E$ and the absolute value of staggered magnetization $|M_{stag}|$ against $p$ with  $\varepsilon$ set to 1.0 for a lattice size $N = 100^2$. Each curve is drawn using symbols (and colour) as indicated in the legend.}
 \label{fig:psmag}
 \end{figure}

To avoid repetition, we present only the phase diagram for the staggered magnetisation as a function of $p$ and $\varepsilon$: Figure~\ref{fig:pspd} shows the absolute value of the staggered magnetization $|M_{stag}|$ of the system at time $t = 512$ for a square lattice of size $N = 100^2$. The paramagnetic region, with low values of $|M_{stag}|$ is coloured black in the figure, whereas the frozen regions (containing either parallel or anti-parallel ordering) with high values of $|M_{stag}|$, are coloured yellow (light grey).
These phases are investigated more quantitatively in  Figure~\ref{fig:psmag}, where we plot the absolute value of  magnetization $|M|$, energy $E$ and  the absolute value of staggered magnetization $|M_{stag}|$ against \textit{p}, with $\varepsilon$ equal to 1.0; each curve is an average over 100 independent runs. The region where both the magnetization $|M|$ and staggered magnetization $|M_{stag}|$ curves saturate to 1, corresponds to parallel alignment, whereas $|M| \approx  0$ with  $|M_{stag}| \approx 1 $  implies an anti-parallel alignment of the spin types. The energy graph is consistent with this interpretation, given the definition of the energy in Equation~\ref{defs}: zero in the PFP, middling in the paramagnetic phase and unity in the AFP.

\begin{figure}[h]
\vspace*{0.44cm}
\centering\includegraphics[scale=0.28]{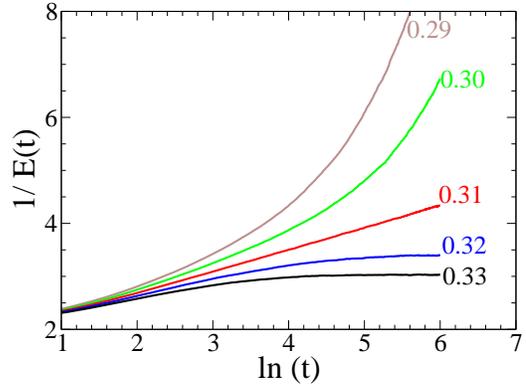}
\caption{(color online)[\textit{ps} update] Plot of  the inverse energy $1/E(t)$ versus  ln $t$ for different values of $p$ close to $p_{c1} = 0.31$,  with  $\varepsilon$ set to 1.0 for a lattice size $N = 256^2$. The $p$-values are indicated on the curves. The straight line corresponding to $p_{c1} = 0.31$ (shown in red) has a slope of approximately $4/3\pi$.}
\label{fig:psv1}
\end{figure}

\begin{figure}[h]
\vspace*{0.4cm}
\centering\includegraphics[scale=0.28]{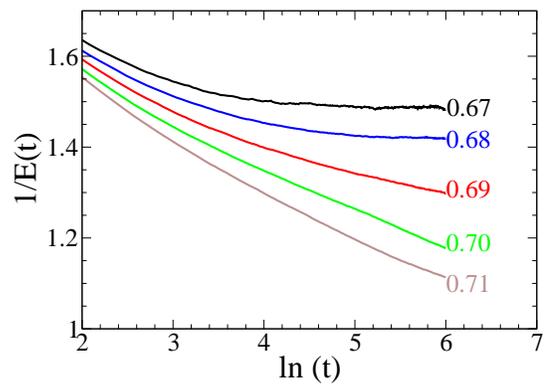}  
\caption{(color online)[\textit{ps} update] Plot of  the inverse energy $1/E(t)$ versus ln $t$ for different values of $p$ close to $p_{c2} = 0.69$,   with  $\varepsilon$ set to 1.0, for lattice size $N = 256^2$. The $p$-values are indicated on the curves. The straight line corresponding to $p_{c2} = 0.69$ (shown in red) has slope $-4/15\pi$ approximately.}
\label{fig:psv2}
\end{figure}

Finally, we present the variation of inverse energy with the natural logarithm of time, ln $t$, near the critical points $p_{c1}$ and $p_{c2}$ in Figure~\ref{fig:psv1} and Figure~\ref{fig:psv2} respectively. 
 Each of the curves is an average over 200 independent runs. At criticality, both plots show a linear proportionality between 1/$E(t)$ and ln $t$, with slopes of $4/3\pi$ and $-4/15\pi$ at $p_{c1}= 0.31$ and $p_{c2}= 0.69$ respectively. Again, this indicates that the \textit{ps} update of the model  belongs to the generalised, rather than the exact, universality class of the voter model \cite{haye}.
 
To conclude, the \textit{ps} update yields qualitatively similar results to the \textit{pp} update, with the appearance of two frozen phases PFP and AFP. Again, small values of $\varepsilon$ indicating longer memories of outcomes, lead to a smearing out of the phase transition, because of the decreasing effectiveness of the outcome-based rule.
% % % % % % % % % % % % % % % % % % % % % % % % % % % % % %
% % % % % % % % % % % % % % % % % % % % % % % % % % % % % %
\subsection{Explanation for the nature of the phase diagrams for different updates}
\label{expl}
In this subsection, we give arguments for the three most important features of the phase diagrams presented above:

\begin{enumerate}[(i)]
\item The appearance of anti-parallel ordering in both  $pp$ and $ps$ updates
\item The symmetry of the PFP and the AFP phases in both $pp$ and $ps$ updates
\item The positioning of the disordered phase in $ss$, $pp$ and $ps$ updates
\end{enumerate}

\begin{figure}[h]
\vspace*{0.4cm}
\centering\includegraphics[scale=0.28]{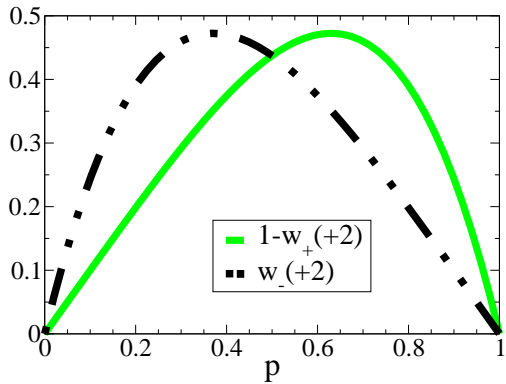}
\caption{(color online) Transition probabilities  ($1-w_{+}(+2)$) (drawn as solid line (in green (grey))) and $w_{-}(+2)$ (drawn as dashed line (in black)) against $p$ [from Equation~\ref{teqns}].}
\label{fig:tp1}
\end{figure}

The clue which explains all of the above, is  the formation of `active' or disparate bonds by the rules of the model under different updates: these are clearly the units of anti-parallel ordering. Consider thus configurations where a site is surrounded by a majority of its own kind: this would correspond to a local field of $+2$ for a $+$, and $-2$ for a $-$. Here the majority of the bonds are `like' or `inactive'. The transition probability for the increase of active bonds from such configurations is $1-w_{+}(+2)$  (or $1-w_{-}(-2)$) [see Equation~\ref{teqns}]. The transition probabilities for the \textit{decrease} of active bonds are given by an opposite scenario, yielding $w_{-}(+2)$ (or $w_{+}(-2)$) [see Equation~\ref{teqns}]. We plot two of these transition probabilities in Figure~\ref{fig:tp1}, corresponding respectively to an increase and a decrease of active bonds: the former peaks at $p=0.63$ while the latter  peaks at $p=0.37$.

\begin{figure}[h]
\vspace*{0.4cm}
\centering\includegraphics[scale=0.28]{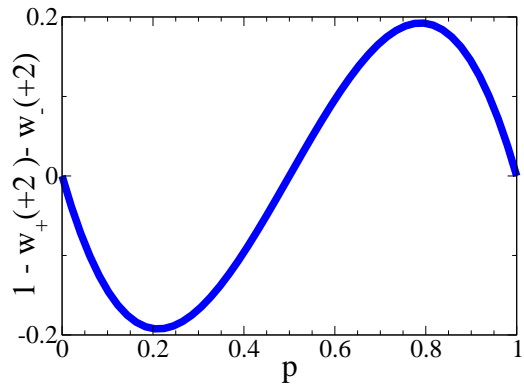}
\caption{(color online) The difference in the transition probabilities ($1-w_{+}(+2)$) and $w_{-}(+2)$ is plotted against $p$ [see Equation~\ref{teqns}].}
\label{fig:tp2}
\end{figure}

The net probability of having active bonds is the difference between these two transition probabilities, and is plotted in  Figure~\ref{fig:tp2}. We see from this that the probability of having active bonds is greatest at
$p = 0.79$, and least at $p = 0.21$. The last ingredient that we need to explain the AFP phase in the $pp$ and $ps$ updates is the fact that once clusters with many active bonds, i.e. anti-parallel ordering, are formed, the majority rule applied via the \textit{parallel} update \textit{preserves} such ordering. With all this in place we see that as expected, the AFP phase in both $pp$ and $ps$ updates shows up in qualitatively the same regions as predicted by Figure~\ref{fig:tp2}, with a peak, in both cases at around $p=0.79$. Correspondingly, the PFP in both $pp$ and $ps$ updates shows up in the region predicted in this figure, with a peak in both cases at around $p=0.21$. Notice (Figure~\ref{fig:tp2}) that the peak and the dip in the probability of active bonds are symmetric about $p = 0.5$, thus explaining the symmetry that we have observed in Figure~\ref{fig:pppd} and Figure~\ref{fig:pspd}; $p = 0.5$ is thus the natural point for the appearance of the disordered phase in both $pp$ and $ps$ updates, as will be confirmed by an inspection of Figure~\ref{fig:pppd}, Figure~\ref{fig:pspd} and Figure~\ref{fig:tp2}.

The only remaining point to be explained is the appearance of the disordered phase in the $ss$ update. In this case too, the analysis leading to Figure~\ref{fig:tp2} for the probabilities of having active bonds remains valid. However, the sequential update of the majority rule always favours strictly parallel ordering, so that typically clusters of active bonds are destroyed once formed. When the probability of their formation is strongest, i.e. at $p=0.63$ (see Figure~\ref{fig:tp1}), the competition between the majority and outcome-based rules is at its most intense, and a disordered phase may be expected to appear. Indeed, the mid-point of the disordered phase for the $ss$ update is shown in Figure~\ref{fig:sspd} to be in exact agreement with this predicted peak, given as it is by $(p_{c1}+p_{c2})/2 = 0.63$.
%%%%%%%%%%%%%%%%%%%%%%%%%%%%%%%%%%%%%%%%%%%%%%%%%%
%%%%%%%%%%%%%%%%%%%%%%%%%%%%%%%%%%%%%%%%%%%%%%%%%%%
 \subsection{The sp update}
 \label{spsection}
\begin{figure}[h]
\centering\includegraphics[scale = 0.37]{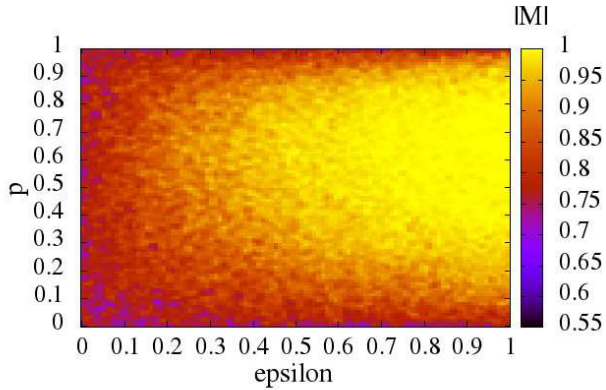} 
\caption{(color online)[\textit{sp} update] Phase diagram of the sp-updated model. Plot of the absolute value of magnetization $|M|$ at time $t = 512$ for lattice size $N = 64^2$ in the p-$\varepsilon$ plane. No phase transition is visible.}
\label{fig:sppd}
\end{figure}

\begin{figure}[h]
\vspace*{0.4cm}
\centering\includegraphics[scale = 0.28]{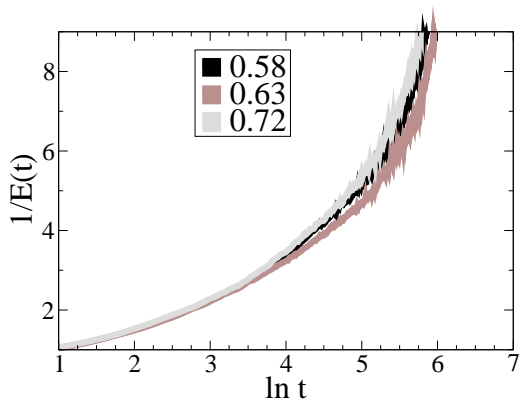}  
\caption{(color online)[\textit{sp} update] Plot of   the inverse energy $1/E(t)$ versus  ln $(t)$ for different values of $p$ and $\varepsilon = 1$, for a lattice size $N = 256^2$. The value of $p$ for each curve is given by a different colour as indicated. No phase transition is visible.}
\label{fig:spEvT}
\end{figure}

In the case of this update, the phase diagram, Figure~\ref{fig:sppd}, shows nothing but a frozen phase. As is evident from the plot of inverse energy 1/$E(t)$ versus  ln $t$ (Figure~\ref{fig:spEvT}), there is a continuous increase in 1/$E(t)$ for all values of $p$ at \(\varepsilon= 1.0\) (where the phase transition is expected to  be the most visible).
This suggests that the two rules, majority and performance-based, do not compete with each other at all (this is what had led to the appearance of the disordered phase in all the other updates). We suggest that this might be because the sequential update (with its more immediate conversions) in the case of the majority rule completely dominates the slower parallel update for the outcome-based rule: this in turn leads to an increasing tendency for consensus, independent of the value of $p$, with which our results are consistent.
% % % % % % % % % % % % % % % % % % % % % % % % % % % % % % % % % % % % % % % % % % % % % % % %
% % % % % % % % % % % % % % % % % % % % % % % % % % % % % % % % % % % % % % % % % % % % % % % % %%%
\section{Away from coexistence: when the strategies are distinct}
\label{away}
Evidently, the real use of a competitive learning model such as this one is when the agents have a choice of distinct strategies. The full exploration of the behaviour of the model at coexistence as carried out in this paper as well as in earlier work \cite{mehta,mahajan} was aimed at an understanding of its phase diagram. However, in the exploration of the behaviour of the model away from coexistence, we hope to gain an understanding of the relative importance of parameters such as superiority of strategy (modelled by $p$) and  memory (modelled by $\varepsilon$), when these are in competition.
The behaviour in asymmetric conditions (using $p_{+} > p_{-}$ and $\varepsilon_{+} > \varepsilon_{-}$)
is formulated in terms of
 the application of two biasing `fields' \cite{mehta}
\begin{equation}
H = p_{+} - p_{-}, \hspace*{1 cm} B=  \varepsilon_{+} - \varepsilon_{-},
\label{hbdef}
\end{equation}  
such that one strategy is favoured over the other. 

In the following subsection, we look at a linear response formulation of our question in terms of unequal $p$'s, viewed as a biasing field, keeping $\varepsilon$ the same for both strategies. In the final subsection, we look at unequal strategies \textit{as well as} unequal memories, to find out whether inferior strategies applied with a good memory of past outcomes, can win overall.
% % % % % % % % % % % % % % % % % % % % % % % % % % % % %
% % % % % % % % % % % % % % % % % % % % % % % % % % % % % %
\subsection{Linear response theory: strategies with unequal $p$}
\label{linsection}
Linear response theory is premised on the basis that an order parameter such as the magnetisation undergoes a sharp change in the neighbourhood of a critical point. In both the $ss$ and $pp$ updates of this model, there are two critical points $p_{c1}$ and $p_{c2}$ separating a paramagnetic phase from two frozen phases. In this subsection, we look at the linear response behaviour of the model in the vicinity of both critical points, starting from the disordered phase: clearly the response will depend both on the value of $p$ as well as on the value of the biasing field $H$ (defined in terms of the difference of the $p$'s in Equation~\ref{hbdef}). In the following, we examine the response by choosing a given value of $p$, and writing $p_{\pm} = p \pm H/2$, keeping $\varepsilon$ fixed.
 \begin{figure}[h]
 \vspace*{0.45cm}
 \centering\includegraphics[scale=0.28]{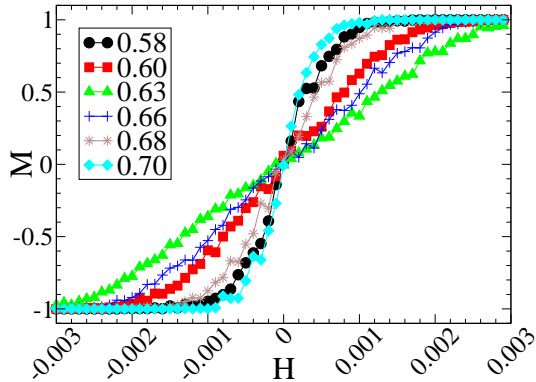}  
 \caption{(color online)[\textit{ss} update] Plot of magnetization $M$ against biasing field $H$ for different values of $p$. Each curve is drawn using different symbols (and colour) as shown in the legend, at time $t = 2000$ with  $N = 100^2$ and $\varepsilon = 1.0$.}
 \label{fig:ssmVh}
 \end{figure}
 
\begin{figure}[h]
 \vspace*{0.4cm}
 \centering\includegraphics[scale=0.28]{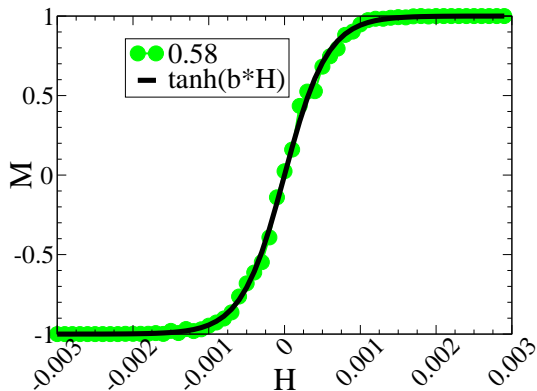}
 \caption{(color online)[ss update] Plot of magnetization $M$ against field $H$ for $p = 0.58$ and $\varepsilon = 1.0$  at time $t = 2000$ with  $N = 100^2$. The fit of a $\tanh(bH)$ curve almost completely overlaps with our numerical results.}
 \label{fig:lrt2}
 \end{figure}

We first consider the $ss$-updated model. Figure~\ref{fig:ssmVh} shows a plot for magnetization $M$ against the biasing field \textit{H} at various values of \textit{p}, that are within the paramagnetic phase at $\varepsilon = 1.0$. Each curve is obtained after averaging over 100 initial configurations using a square lattice of size $N = 100^2$. For each \textit{p} in the paramagnetic phase, we see a linear behaviour of $M$ against $H$ around $H \sim 0$, with  all subsequent increases in the field strength leading to saturation, as expected. 
 For a given $p$ value we observe a functional dependence of the form
\[M = \tanh(bH)\]
where \[b \propto (p_{central}+p)^2\]
taking \[p_{central} = (p_{c1}+p_{c2})/2. \]
The quality of the fit to $\tanh(bH) $ is seen Figure~\ref{fig:lrt2}: the black fitting curve almost completely coincides with a sample curve taken from  Figure~\ref{fig:ssmVh}.
 
\begin{figure}[h]
 \centering\includegraphics[scale=0.37]{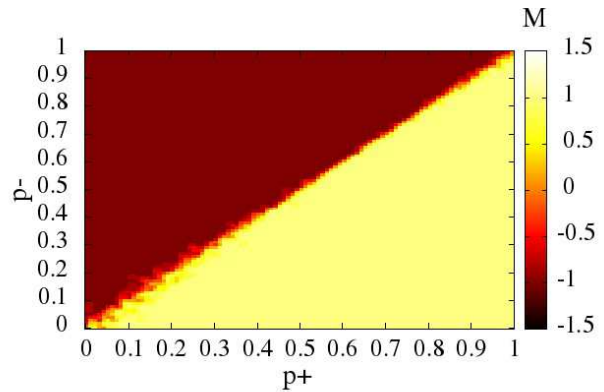}
 \caption{(color online)[\textit{ss} update] Plot of magnetization $M$ in the $p_{+}$-$p_{-}$ plane at time $t = 7000$, for a lattice of size $N = 100^2$ with $\varepsilon = 1.0$. The  $+$ strategy dominates in the yellow (light grey) region, while the $-$ strategy dominates in the brown (black) region. }
 \label{fig:ssmagpp}
 \end{figure}

These results also admit of an alternative representation, shown in Figure~\ref{fig:ssmagpp}, where it is clear that the relative values of the bias correspond to different regions of domination of each strategy in phase space.
 
\begin{figure}[h]
\vspace*{0.44cm}
\centering\includegraphics[scale=0.28]{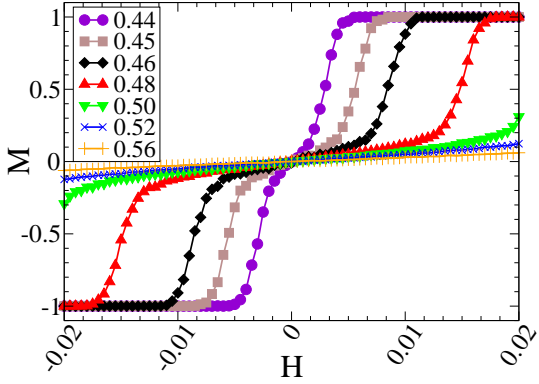}  
\caption{(color online)[\textit{pp} update] Plot of magnetization $M$ against field $H$ for different values of $p$, each curve indicated by a different symbol (and colour) as shown in the legend, at time $t = 2000$ with  $N = 100^2$ and $\varepsilon = 1.0$.}
\label{fig:ppmVh}
\end{figure}

We next examine the linear response behaviour  of the $pp$-updated model, again in the vicinity of the two critical points. Figure~\ref{fig:ppmVh} is a plot showing the variation in magnetization $M$ along the field $H$ for different $p$ values at $\varepsilon = 1.0$. For the lower values of $p$, in the vicinity of $p_{c1}$, we see 
very similar behaviour to that presented in Figure~\ref{fig:ssmVh}, corresponding to an expected $\tanh(bH)$ behaviour as shown in Figure~\ref{fig:lrt2}: the PFP phase lying to the left of $p_{c1}$ is, after all, identical to the frozen phases in the $ss$ update. As we approach the vicinity of $p_{c2}$, the curves are markedly different: the nature of the ordered phase is one that corresponds to magnetisation values of 0 (see orange curve drawn using plus symbols in Figure~\ref{fig:ppmVh}), which is again consistent with the AFP phase that lies to the right of $p_{c2}$. 
 
\begin{figure}[h]
\vspace*{0.45cm}
\centering\includegraphics[scale=0.28]{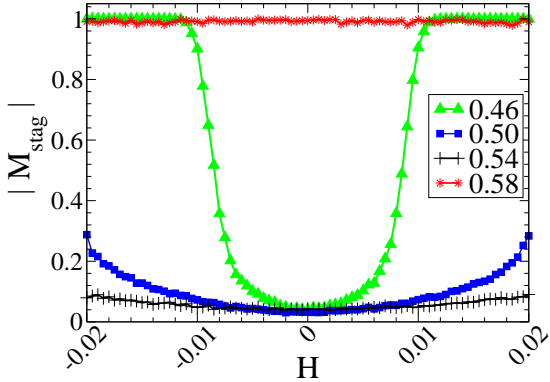}  
\caption{(color online)[\textit{pp} update] Plot of the absolute value of staggered magnetization $|M_{stag}|$ against field $H$ for different values of $p$, each curve indicated by a different symbol (and colour) as shown in the legend, at time $t = 2000$ with  $N = 64^2$ and $\varepsilon = 1.0$.}
\label{fig:ppmsVh}
\end{figure}

To establish this more firmly we look at plots of  the absolute value of the staggered magnetisation $|M_{stag}|$ as a function of bias $H$, in Figure~\ref{fig:ppmsVh}. The green (triangle), blue (square) and black (plus) lines denote increasing values of $p < p_{c2}$, where the staggered magnetisation increasingly approaches zero, as expected in the disordered phase: however the red (star) line, corresponding to $p > p_{c2}$ shows an abrupt jump in the value of $|M_{stag}|$ to unity. Combined with the analysis of the previous paragraph, this shows convincingly that the phase we refer to as AFP indeed corresponds to anti-parallel ordering.
  
\begin{figure}[h]
  \centering\includegraphics[scale=0.37]{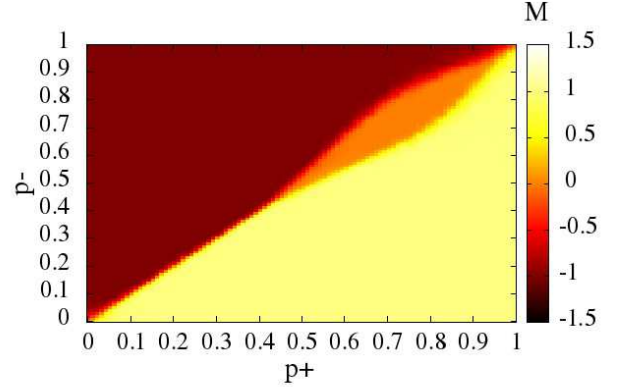}
  \caption{(color online)[\textit{pp} update] Plot of magnetization $M$ in the $p_{+}$-$p_{-}$ plane for a lattice of size $N = 100^2$ at time $t = 7000$, with      $\varepsilon = 1.0$. As before, the regions of $+$ and $-$ strategy domination are coloured yellow (light grey) and brown (black); the orange (grey) region corresponds to \textit{both} the paramagnetic and the AFP region (see text). }
  \label{fig:ppmagpp2}
  \end{figure}

\begin{figure}[h]
    \centering\includegraphics[scale = 0.37]{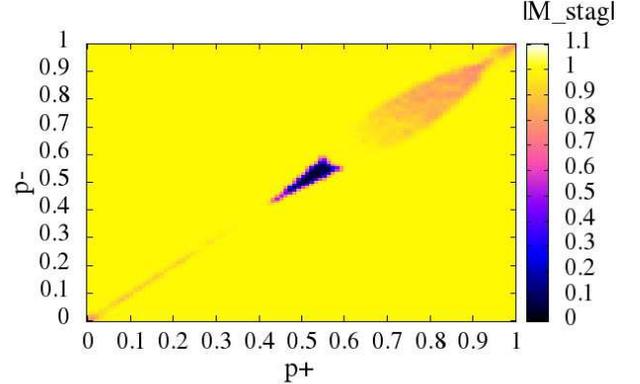}
    \caption{(color online)[\textit{pp} update] Plot of staggered magnetization $|M_{stag}|$ in the $p_{+}$-$p_{-}$ plane,  for a lattice of size $N = 100^2$ at time $t = 7000$,  with $\varepsilon = 1.0$. Here, yellow (light grey) represents the region of parallel ordering, black represents the disordered phase, and the light brown (grey) represents AFP order.}
    \label{fig:ppmagpp1}
    \end{figure}

We present below an alternative representation of the above results for ease of visualisation.  In Figure~\ref{fig:ppmagpp2}, the magnetisation $M$ is plotted in the $p_{+}$-$p_{-}$ plane: as before, the regions of brown (black) (resp. yellow (light grey)) correspond to domination by $-$ strategies (resp. $+$ strategies). Notice, however, that the coexistence line has an island of very low magnetisation: in actual fact, this corresponds to the regions of \textit{both} the paramagnetic and AFP phase. This is clearer in the plot of the absolute value of the  staggered magnetisation $|M_{stag}|$, shown in Figure~\ref{fig:ppmagpp1}, where the black portion of the island along the coexistence line corresponds to the disordered phase, while the faintly brown (grey) portion corresponds to the AFP.
  
 These plots allow us to go beyond the previous analysis in defining the domain of stability of the AFP phase: we see clearly from Figure~\ref{fig:ppmagpp2} and Figure~\ref{fig:ppmagpp1} that the AFP phase exists for $p > p_{c2}$ \textit{only if} the biasing field is within the bounds defined by $H=|$$p_{+} - p_{-}$$|$ $\leq 0.19 \pm 0.02$. In qualitative terms, this implies that at least in the absence of memory, when the two strategies have nearly equal success rates, neighbouring agents may adopt different strategies~\cite{zhao} in equilibrium. % % % % % % % % % % % % % % % % % % % % % % % % % % % % % % % % % % % % % % % 

Having thoroughly investigated the linear response regime for the $ss$- and $pp$-updated models, we will now examine the effect of the memory parameter $\varepsilon$ in the next subsection.
 % % % % % % % % % % % % % % % % % % % % % % % % % % % % % % % %
 % % % % % % % % % % % % % % % % % % % % % % % % % % % % % % % % %
\subsection{Role of memory parameters: the case of unequal $\varepsilon$}
\label{rolesection}
The principal competition in this model is that between two strategies with different global success rates $p$, which determines the relative dominance of each one in phase space. The memory parameter $\varepsilon$ plays a more subtle role in this competition: although it cannot be a determinant of phase behaviour in the way that the success rates are (as a consequence of the rules elucidated in Equation~\ref{teqns}), it \textit{can}, as we will show, cause a surprising change in the dominance of an ostensibly superior strategy. In \cite{mehta}, it had been suggested that agents with inferior strategies and good memories might indeed win against agents who had better strategies but worse memories. Here, we will make this prediction more quantitative.

\begin{figure}[h]
 \centering\includegraphics[scale=0.30]{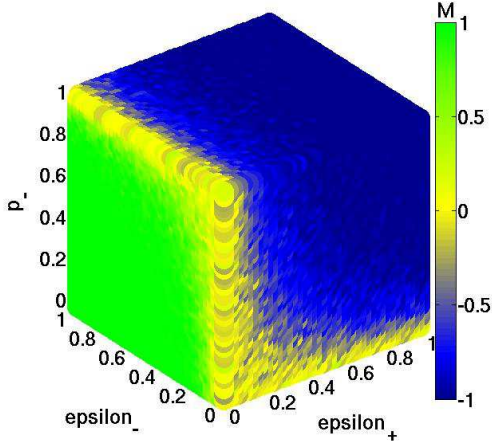}
 \caption{(color online)[pp update] A $3D$ plot of Magnetization $M$ with parameters $\varepsilon_{+}$ (along $x$), $\varepsilon_{-}$ (along $y$) and $p_{-}$ (along $z$), setting $p_{+} = 0.50$ for a lattice of size $N = 64^2$ at time $T = 2000$. Green (grey) denotes the dominance of the $+$'s, while blue (black) denotes that of the $-$'s. The other colours represent cases of intermediate ordering.}
 \label{fig:pp3d1}
 \end{figure}

The phase diagram of the model away from coexistence involves four parameters, $p_{\pm}, \varepsilon_{\pm}$, so that its representation is a non-trivial problem. In the following, we choose to fix $p_{+}$ to 0.5, and to vary the other three parameters: a sample $3d$ plot is shown in Figure~\ref{fig:pp3d1}. We analyse the three visible faces in detail, before remarking on the phase behaviour within the cube: the colour coding is such that green (grey) represents dominance of $+$ strategies, blue (black) represents dominance of $-$ strategies, and other colours represent mixed states.
\begin{itemize}
\item The leftmost face of the cube corresponds to the plane $\varepsilon_{+} = 0$; this implies that the agents using $+$ strategies will never convert, no matter what the outcome-based rule says. The minimum occupancy of $+$ strategies for random configurations should thus be of the order of $N/2$, which can only increase depending on the conversions of agents using $-$ strategies into the camp of the $+$'s. The bottom line corresponds to $p_{-} = 0$, which is when such conversions are maximal (so that all $N$ sites are $+$): the green (grey) colour is at its most pronounced here, changing gradually over to other colours only as $\varepsilon_{-} \to 0$ to the right of the line, when agents using $-$ strategies too begin to refuse to convert, irrespective of the outcome rules. As the values of $p_{-}$ increase beyond 0.5 (the fixed value for $p_{+})$, we note that the dominance of the $+$ strategy gradually gives way to states with a mixture of strategies; when $\varepsilon_{-} \to 0$, this tendency is at its most pronounced, while when $\varepsilon_{-} \to 1$, this is at its least pronounced, since local conversions can sometimes go against global success rates.
\item The front face of the cube corresponds to $\varepsilon_{-} = 0$; this implies that agents using $-$ strategies will never convert, no matter what the outcome-based rule says. This is a reflection of the previous case, where the minimum number of $-$ sites is once again $N/2$, which can only be increased as the conversions from the $+$'s add to it. 
\item The top face of the cube corresponds to $p_{-}= 1$, where globally a predominance of the $-$ strategy is expected. This is found over almost all the range of  $\varepsilon_{-}$ except at low values of $\varepsilon_{+}$, where agents using the $+$ strategy refuse to convert, despite their globally poorer performance.
\end{itemize}
 \begin{figure}[h]
 \centering\includegraphics[scale=0.32]{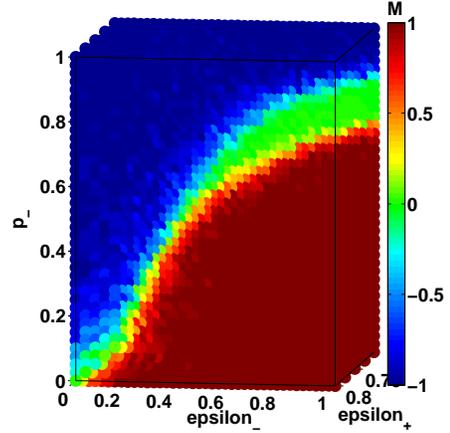}
 \caption{(color online)[pp update] A $3D$ plot of Magnetization $M$ for the parameters $\varepsilon_{+}$ (lying between $0.7$ and $0.8$), $\varepsilon_{-}$ and $p_{-}$, with $p_{+} = 0.70$, for a lattice of size $N = 64^2$ at time $T = 2000$. The red (lower black) region represents the dominance of the $+$ strategy and the blue (upper black) region represents the dominance of the $-$ strategy. The green (grey) region represents AFP.}
 \label{fig:pp3d2}
 \end{figure}
The interior of the cube can show markedly different behaviour, which we illustrate via a sample slice shown in Figure~\ref{fig:pp3d2}. In this figure, red (lower black) and blue (upper black) regions correspond to the dominance of $+$ and $-$ strategies respectively. Here we set the value of $p_{+}$ to 0.7, and look at a slice of its phase space cube, as before: choosing $\varepsilon_{+}$ to be between 0.7 and 0.8, we look at the dominating strategy as a function of
the variables $\varepsilon_{-}$ and $p_{-}$. If the memory parameters had not existed, we would have expected the $-$ strategy (blue (upper black) region in the figure) to predominate only for $p_{-} > 0.7$; the red (lower black) region  would have been covering the entire slice below this, corresponding to the dominance of the $+$ strategy. However, the reality is rather different. The $+$ strategy does indeed predominate for $p_{-} < 0.7$, provided that agents using the $-$ strategy have imperfect memory; but there is a striking predominance of the $-$ strategy (even for very low values of $p_{-}$) provided that the memory of the agents employing this strategy, is much better than those of the other kind ($\varepsilon_{-} \ll \varepsilon_{+}$).
 \begin{figure}[h]
 \centering\includegraphics[scale=0.28]{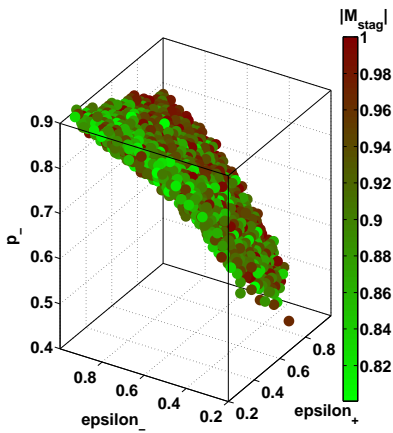}
 \caption{(color online)[pp update] A $3D$ plot of absolute staggered magnetization $|M_{stag}|$ with parameters $\varepsilon_{+}$(along $x$), $\varepsilon_{-}$(along $y$) and $p_{-}$(along $z$) for $p_{+} = 0.50$ and lattice size $N = 64^2$, at time $T = 2000$. The graph delineates the region of AFP for these parameter values.}
 \label{fig:pp3d3}
 \end{figure}
 
 A last feature to mention is the green (grey) region in Figure~\ref{fig:pp3d2}: here, there is a region of alternating $+$ and $-$ ordering (AFP), corresponding to ongoing competition between the two strategies. This phenomenon is most pronounced when the two strategies are equally successful, and are  \textit{both} accompanied by weak memories of earlier outcomes. In Figure~\ref{fig:pp3d3}, the structure of the full AFP is shown (by selecting phase points with low values of absolute magnetisation $|M|$ and high absolute staggered magnetization $|M_{stag}|$) as a function of $p_{-}$ and $\varepsilon_{\pm}$, fixing $p_{+} = 0.5$.
 % % % % % % % % % % % % % % % % % % % % % % % % % % % % % % % % % % %
 % % % % % % % % % % % % % % % % % % % % % % % % % % % % % % %
\section{Discussion}
\label{discuss}
The work of this paper extends work done on a problem of strategic learning \cite{mehta,mahajan} which, although originally suggested by a problem on technology diffusion \cite{kalyan}, has much wider ramifications (e.g., in relation to threshold learning dynamics \cite{vega}). 

In any agent-based modelling scheme, it is important to know whether agents react sequentially or collectively to the spread of information. Our results show that these issues make a quantitative as well as a qualitative difference to the results, changing not just exponents but also the entire nature of the phase diagram in most cases. Given that, typically, the propagation of technologies through well-connected societies is of interest~\cite{kalyan}, we choose ordered rather than random updates, and examine the response of the model of~\cite{mehta} to all possible combinations of sequential and parallel updating. From the viewpoint of theoretical physics, a major result is that this model is robustly in the universality class of the voter model~\cite{haye}, for all but one of the updates. This strong relationship with the voter model results from the model of~\cite{mehta,mahajan} being driven by interfacial noise alone, i.e. the absence of surface tension~\cite{haye}.

Another major result, still to do with updates, is the appearance of a phase of anti-parallel ordering (AFP) in the high-performing limits of $p$, for both the $pp$ and the $ps$ updates. While the technicalities behind this are explained in the text, we give here a more intuitive reason for this, from the perspective of strategic learning. The parallel scheme can be viewed as a more `equilibrated' update than the sequential one, since it gives a chance for the entire lattice to be updated `simultaneously'. It is then natural that in the regime that both agents are high-performing, they should be equally preferred: this lies behind the `alternating' order inherent in the AFP regime. By contrast, since the sequential paradigm corresponds to a `non-equilibrium' update, where every agent responds to the updated value of its neighbours, the above logic leads to a disordered phase where every prescription of the outcome-based rule is countermanded by the following majority rule. Using once again the illustration of propagating technologies \cite{kalyan}: when all the populace have equal and \textit{simultaneous} access to information about two high-performing technologies, we will see the coexistence of both \cite{zhao} (as predicted by the AFP phase), whereas when information about each one is passed on \textit{sequentially}, the conflicting information so obtained can result in sheer disorder. Finally, we mention here that our investigation of different updates on this game-theoretic model has been applied to related game-theoretic models of cognitive learning and synaptic plasticity \cite{epl, plos}, where updates relate to the directionality of synapses in a network.

Moving away from the domain of critical behaviour at coexistence, we have looked at the behaviour of the competitive learning model when the two strategies have distinct attributes (this, after all, is truer to the title of competitive learning!). To begin with, we have examined the response of the model to unequal strategies $p_{\pm}$, and have found in general that the smarter strategy wins  (for equal values of the memory parameter $\varepsilon$), as might be expected. An interesting feature is that the region of anti-parallel ordering (AFP) found earlier still persists in the presence of bias, provided that the difference in $p$ is below a well-defined bound: in other words, when two distinct strategies are \textit{almost} equally successful, one will typically find that they can coexist in society. Finally, we have looked at the effect of memory:  we have found that while memory has a secondary role in determining the phase behaviour of the model, it has a particularly striking effect in turning around the results of any bias in $p$. A major result of our paper is thus that decisions based on a good memory of earlier outcomes can, within limits, compensate for the choice of inferior strategies.
% % % % % % % % % % % % % % % % % % % % % % % % % % % % % % % % %
% % % % % % % % % % % % % % % % % % % % % % % % % % % % % % % % % 
\begin{acknowledgments}
AAB would like to thank Dr. G. Mahajan and Mr. B. Chakraborty for helpful discussions. AM acknowledges the award of a grant from the DST (Govt. of India) through the project ``Generativity in Cognitive Networks", through which AAB was supported.
\end{acknowledgments}
% % % % % % % % % % % % % % % % % % % % % % % % % % %
% % % % % % % % % % % % % % % % % % % % % % % % % % % % %

\end{document}